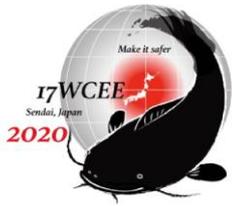

# A NONPARAMETRIC UNSUPERVISED LEARNING APPROACH FOR STRUCTURAL DAMAGE DETECTION

K. Eltouny[(1)], X. Liang[(2)]

[(1)] *Ph.D. Candidate, Department of Civil, Structural and Environmental Engineering, University at Buffalo, the State University of New York, NY, United States, keltouny@buffalo.edu*
[(2)] *Assistant Professor of Research, Department of Civil, Structural and Environmental Engineering, University at Buffalo, the State University of New York, NY, United States, liangx@buffalo.edu*

## *Abstract*

In a world of aging infrastructure, structural health monitoring (SHM) emerges as a major step towards resilient and sustainable societies. The current advancements in machine learning and sensor technology have made SHM a more promising damage detection method than the traditional non-destructive testing methods. SHM using unsupervised learning methods offers an attractive alternative to the more commonly used supervised learning since it only requires data of the structure in normal conditions for the training process. The density-based novelty detection method provides a statistical element to the damage detection process but it relies heavily on the accuracy of the estimated probability density function (PDF). In this study, a novel unsupervised learning approach for SHM is proposed. It is based on the Kernel Density Maximum Entropy method by leveraging Bayesian optimization for hyperparameter tuning and also by extending the method into the multivariate space by the use of independent components analysis. The proposed approach is evaluated on a numerically simulated three-story reinforced concrete moment frame, where 94% of accuracy is achieved in structural damage detection.

## 1. Introduction

The ability to recover in a timely manner after natural hazards is one of the greatest characteristics of resilient buildings. This is possible if proper damage detection methods are employed. Vision-based (e.g., using drones [53-56]) and non-destructive evaluation methods are commonly used for damage detection. However, these methods require expensive equipment. Additionally, the approximate damage location must be known as a priori resulting in a difficult damage evaluation process for internal non-exposed members. Data-driven structural health monitoring (SHM) provides an attractive alternative to assessing structural integrity. It utilizes statistical learning techniques with damage-sensitive features extracted from structure measurement data. In the vibration-based SHM, such data is typically acceleration time-series.

The statistical learning approach can be supervised or unsupervised. Supervised learning algorithms, in the context of SHM, use both undamaged and damaged conditions data for training. Many SHM algorithms using supervised learning have been proposed, including using artificial neural networks [1, 2], support vector machines [3-8], convolutional neural networks [9, 10, 47-50], decision trees, and random forests [11]. However, damaged condition data needs to be acquired through numerical simulations or experiments. Accurate finite element models can be very difficult for complex structures while experiments may have physical limitations and are often performed on structural components or scaled specimens. Unsupervised learning serves as a good alternative to supervised learning, especially in the context of SHM. It is trained using only data from the structure in the undamaged condition, and that is usually available in abundance. Novelty detection approaches are usually utilized for unsupervised learning SHM, including Mahalanobis squared distance-based outlier analysis [12], kernel density estimation (KDE), auto-associative neural networks (AANN) [13, 14], self-organizing neural networks [15], and deep restricted Boltzmann machine [16]. Clustering algorithms were also used for SHM, including k-means clustering [17, 18], Gaussian mixture models combined with Genetic Algorithms (GA) for parameter initialization [19], concentric hyperspheres combined with GA for locating clusters decision boundaries [20], and density peaks-based fast clustering algorithm [21].

In previous studies, it was found that outlier analysis performed relatively well only if the data follows a Gaussian distribution, while KDE suffers from the "curse of dimensionality" and is susceptible to overfitting. Kernel Density Maximum Entropy (KDME) method, a non-parametric method, can be leveraged to reconstruct the true PDF[22, 23]. It is computationally efficient compared to other Maximum Entropy (ME) methods and also provides a more accurate estimation on the distribution tails with less amount of training data. However, the convexity of the objective function used for optimizing the coefficients is not guaranteed.

In this paper, we propose a robust SHM framework capable of detecting damage solely based on normal condition data. The framework is based on Bayesian-optimized KDME for novelty detection using cumulative intensity measures as damage-sensitive features. A 3D finite element model of a three-story reinforced concrete (RC) moment frame, subjected to 100 different earthquake time histories with incorporated environmental variation effects, is used to assess the performance of the proposed framework. The remainder of the paper is organized as follows. In section 2, the damage-sensitive feature extraction process is described. Section 3 introduces the proposed Bayesian-optimized KDME framework along with the necessary formulations. Section 4 presents the numerical case study along with its results. Finally, conclusions and summary are presented.

## 2. Damage-sensitive feature extraction

Damage-sensitive feature selection is one of the most critical elements in developing SHM systems. A recently developed damage-sensitive feature, Cumulative Absolute Velocity (CAV), has been shown to have a good correlation with earthquake-induced damage [6, 24, 25]. The cumulative intensity measure ($I^\eta$) is a general form of CAV, given as follows:

$$I^{\eta} = \int_0^T |a|^{\eta} dt \tag{1}$$

where $a$ is the absolute acceleration value at time-step $t$, $T$ is the total duration of the particular acceleration time-series segment, and $\eta$ is a hyperparameter. It is a low-dimensional damage-sensitive feature extracted from acceleration time-series that has been used to identify and locate damage in various SHM applications [7, 8, 10]. In a single-degree-of-freedom system, $I^{\eta}$ can be estimated at both the input (source) and output (response) locations and the two together contain information about the input and output energy and can reflect dynamical properties from the normal condition of this system. In a shear building, this feature can be exploited for damage detection by using $I^{\eta}$ at the top floor (output) and the base (input). However, the hyperparameter $\eta$ requires data for structures in both normal and damaged cases to be optimized. This is generally not available for unsupervised learning. Instead, a set of $I^{\eta}$ with different $\eta$ values is used as a feature vector. In this study, the set of $\eta$ is bounded between 0.1 and 10. The domain is discretized using a step of 0.1 thus creating a set of 100 different values of $\eta$. Additionally, there are two cumulative intensity measures of concern, the first is estimated at the top floor ($I_t^{\eta}$) while the second is estimated at the base ($I_g^{\eta}$). A minimum of 4 accelerometers are required, two for each location corresponding to the two horizontal directions. The final stacked feature vector is given as follows:

$$X = [\underbrace{I_{g,1}^{\eta_1}, \cdots, I_{g,1}^{\eta_{100}}, I_{t,1}^{\eta_1}, \cdots, I_{t,1}^{\eta_{100}}}_{Direction\ 1}, \underbrace{I_{g,2}^{\eta_1}, \cdots, I_{g,2}^{\eta_{100}}, I_{t,2}^{\eta_1}, \cdots, I_{t,2}^{\eta_{100}}}_{Direction\ 2}] \tag{2}$$

Subsequently, PCA is performed on the feature vector $X$ shown in Eq. (2) of the training data. PCA is a linear transformation that is commonly used for dimension reduction. Through PCA, $q$ principal components with the highest variance variables are selected to be the representative damage-sensitive features where $q$ is an integer less than the dimension of $X$. Next, the loadings and means used in the transformation of the training data are stored for transforming new observation into the principal components space.

Independent Component Analysis (ICA) is performed for the principal components resulting in new independent feature components. This allows for the marginal KDME PDF to be estimated for each independent component and the joint PDF is simply the multiplication of all marginal PDFs. ICA is a blind source separation (BSS) technique. It attempts to find original signals from linear combinations of those signals without any information about the original signal. Assuming that the original signals do not follow Gaussian distribution, ICA attempts to find signals that are as far from Gaussian as possible [26]. This is done by attempting to find a mixing matrix (similar to the loadings for PCA) that linearly transforms independent signals to the mixed signals space. The most common way of finding this matrix involves the maximization of non-Gaussianity measures such as kurtosis and negentropy. Kurtosis is the normalized fourth-order cumulant and is zero for Gaussian distributions. For most non-Gaussian distributions, kurtosis is nonzero. In this study, the robust ICA algorithm, which makes use of the kurtosis contrast, is taken advantage of in performing ICA [27]. Consider a source $s$ that needs to be extracted from pre-whitened mixed sources $x$ using the weight vector $w$ such that

$$s = w^T x \tag{3}$$

If the mixed data is pre-whitened, the kurtosis becomes the fourth-order moment and the weight vector updating rule becomes a gradient descent algorithm with a constant learning rate $\alpha$ such that

$$w^+ = w - \alpha \nabla \mathcal{M}(w) \tag{4}$$

The robust ICA attempts to perform an exact line search of the absolute kurtosis with the search direction being the gradient of the kurtosis with respect to the weight vector. The globally optimal step size is found as follows:

$$\alpha_{opt} = arg\ \max_{\alpha} |kurt(w + \alpha \nabla kurt(w))| \tag{5}$$

Zarzoso and Comon [27] provide details on how to compute optimal step size $\alpha_{opt}$ through the roots of a fourth-degree polynomial.

## 3. Bayesian-optimized kernel density maximum entropy novelty detection

### 3.1 Kernel density maximum entropy distribution

This section briefly summarizes the Kernel Density Maximum Entropy (KDME) method. Every probability distribution has a moment sequence that can be obtained. The inverse mapping is called the Moment Problem (MP) and is not straightforward. KDME method reconstructs the true PDF through a superposition of kernel densities multiplied by non-negative coefficients which sum up to one. The coefficients are acquired through the use of a discrete maximum entropy (ME) approach with fractional moments defined as $E[x^\mu] = \sum_{i=1}^{N} x_i{}^\mu \cdot p_i$ for a real number $\gamma$ and $N$ evaluation points. This approach approximates the PDF accurately using a lower number of moments, and requires less computational effort. Besides, the use of fractional moments results in a more accurate estimation for the PDF including better modeling for the tails (an essential part in novelty detection) without the need for higher-order moments [23, 28, 29].

To better illustrate KDME method, consider a discrete random variable $X$ that has a probability distribution given by $p_i$ evaluated at $x_i$, where $i = 1,2,...,N$. The only information available is the $M$ number of fractional moments $E[X^{\gamma_k}]$, $k = 1,2,...,M$. and the normalization condition $\sum_{i=1}^{N} p_i = 1$. To find $p_i$, additional ($N$-2) conditions are needed. However, according to Jaynes [30], one can use the principle of Maximum Entropy to get the least biased estimate of $p_i$ given the available information (constraints). Thus, Shannon's entropy, which is a metric for the entropy of discrete random variables, can be maximized to find the ME probabilities. Using the method of Lagrange multipliers, the ME probabilities is found by:

$$p_i^{ME}(\lambda) = \frac{1}{m_0}\exp\left(-\sum_{k=1}^{M}\lambda_k x_i{}^{\gamma_k}\right) \tag{6}$$

where $i = 1,2,...,N$ and $m_0 = \sum_{i=1}^{N} exp\left(-\sum_{k=1}^{M}\lambda_k x_i{}^{\gamma_k}\right)$ is the normalization coefficient. Lagrange multipliers are optimized by minimizing the Kullback-Leibler (KL) divergence between $\boldsymbol{p}$ (constant) and $\boldsymbol{p}^{ME}$. Fractional moments, however, are only proven to exist for positive random variables [31]. To estimate the PDF using this method for random variables that could take non-positive values, a coordinate transformation to a positive-only space is required. By choosing a window of analysis $\hat{\Omega} \equiv [x_{min}, x_{max}]$ that is large enough so that the target PDF is zero if $x$ is close to infinity, the data is normalized to a bounded interval $\hat{\Omega}_z \equiv [0,1]$ through a min-max coordinate transformation resulting in a new random variable $Z$ bounded by that interval and maps to the original random variable $X$. Finally, the fractional moments which are now defined for $Z$ can be approximated by the sample fractional moment which, for a sample of size $n$, is $E[z^\gamma] \cong \frac{1}{n}\sum_{i=1}^{n} z_i{}^\gamma$.

In addition to optimizing the Lagrange multipliers, the fractional powers $\gamma$ must be tuned. The following system of linear equations provides a way to linearly estimate $\boldsymbol{\lambda}$ parameters given $\gamma$ avoiding a nested optimization and reducing the computational costs [23]:

$$\boldsymbol{\lambda} = [\mathbf{P}(\boldsymbol{\gamma})]^{-1}\boldsymbol{\rho}(\boldsymbol{\gamma}) \tag{7}$$

where

$$P_{jk}(\boldsymbol{\gamma}) = \gamma_k E[Z^{\gamma_k + \gamma_j}] \qquad \rho_j(\boldsymbol{\gamma}) = (\gamma_j + 1)E[Z^{\gamma_j}] \tag{8}$$

with $j = 0,1,...,M-1$; $k = 1,2,...,M$. Therefore, for a given vector $\boldsymbol{\gamma}$, Lagrange multipliers can be linearly obtained from the solution of Eq. (7).

Since this method implements a discrete Maximum Entropy approach, a kernel density representation is employed to result in a smooth continuous PDF [32]. The PDF is expressed as a superposition of Kernel Density Functions (KDFs). In this study, the KDF is chosen to be Gaussian which is suitable for unbounded target distributions. Thus, the kernel representation for $\boldsymbol{p}^{ME}$ provided at evaluation points $x_{eval,i}$ (or $z_{eval,i}$) is as follows:

$$f_{KDME}(x; \boldsymbol{p}) = \sum_{i=1}^{N} p_i^{ME}K(x; x_{eval,i}, h) = \sum_{i=1}^{N} p_i^{ME}\frac{1}{h\sqrt{2\pi}}\exp\left\{-\frac{1}{2}\left(\frac{x - x_{eval,i}}{h}\right)^2\right\} \tag{9}$$

where $K(x; x_i, h)$ are the KDFs centered at $x_{eval,i}$ (location parameter) belonging to the sample space of the random variable $X$. $h$ is the smoothing parameter or bandwidth. $p_i^{ME}$ is the ME probability at point $x_{eval,i}$. To specify evaluation points $x_{eval,i}$, a practical way is to define a constant step $\Delta x = x_{eval,i+1} - x_{eval,i}$, $i = 1,2,\ldots,N-1$. $N$ must be chosen as a high number such that $h \to 0$ and $K(z;z_i,h) \to \delta(z-z_i)$ and thus, the choice of the KDF does not significantly affect the PDF estimation.

The objective function directly tests the KL-divergence between the target PDF in $Z$ coordinates $f_Z(z)$ and its KDME approximation $f_{KDME}(z;M,\lambda,\gamma)$, this results in the following objective function:

$$\Theta(M,\boldsymbol{\gamma}) = -\frac{1}{n}\sum_{j=1}^{n}\log\left[f_{KDME}(z_j;M,\boldsymbol{\gamma})\right] + \frac{M}{n} \tag{10}$$

where $f_{KDME}(z_j;M,\boldsymbol{\gamma})$ is the kernel representation of $\boldsymbol{p}^{ME}$ in $Z$ coordinates given in Eq. (9), $M$ is the number of moments considered, and $n$ is the sample size. The objective function is equivalent to a modified negative logarithmic likelihood function of $f_{KDME}$ and the final term is added to discourage model complexity. $\boldsymbol{\lambda}$ was dropped because it is estimated linearly for any given $\boldsymbol{\gamma}$ using Eq. (7). Finally, the parameters can be obtained through the minimization of $\Theta(M,\boldsymbol{\gamma})$ as follows

$$\left(M_{opt},\boldsymbol{\gamma}_{opt}\right) = \max_{M}\left\{\max_{\boldsymbol{\alpha}}\{\Theta(M,\boldsymbol{\gamma})\}\right\} \tag{11}$$

## 3.2 Hyperparameters optimization

The convexity of the objective function $\Theta(M,\boldsymbol{\gamma})$ in $\boldsymbol{\gamma}$ is not guaranteed. For this reason, Bayesian optimization is leveraged in optimizing $\boldsymbol{\gamma}$ values. It is a robust technique for attempting to find a global minimum with relatively few iterations. This is done through additional computations that determine the next point to evaluate through incorporating prior belief about the objective function [34]. Although there are multiple models that can be used as a prior distribution, the Gaussian Process (GP) has been commonly used for this task [35].

The optimization task here is to find the global minimum of $\Theta(M,\boldsymbol{\gamma})$ given in Eq. (10) by changing the variables $\boldsymbol{\gamma} = \gamma_1,\gamma_2,\ldots,\gamma_M$ for $M$ number of fractional moments. A domain for $\boldsymbol{\gamma}$ is chosen such that $\gamma_k \in [0, \gamma_{max}]$. For an initial seed of randomly generated vectors $\boldsymbol{\gamma}_1,\boldsymbol{\gamma}_2,\ldots,\boldsymbol{\gamma}_t$, their corresponding objective function estimates $\Theta(\boldsymbol{\gamma}_1),\Theta(\boldsymbol{\gamma}_2),\ldots,\Theta(\boldsymbol{\gamma}_t)$ have a joint multivariate Gaussian distribution such that

$$\boldsymbol{\Theta}_{1:t} \sim \mathcal{N}(0,\mathbf{K}) \tag{12}$$

where $\boldsymbol{\Theta}_{1:t} = \Theta(\boldsymbol{\gamma}_{1:t})$ and $\mathbf{K}$ is the kernel matrix containing the kernels $\kappa(\boldsymbol{\gamma}_i,\boldsymbol{\gamma}_j)$ as its elements. The kernel determines the smoothness properties of samples drawn from the GP. The kernel employed in this study is the Automatic Relevance Determination (ARD) Matérn 5/2 kernel [36]. Given the observations from the previous distribution $\mathbf{D}_{1:t} = \{\boldsymbol{\gamma}_{1:t},\boldsymbol{\Theta}_{1:t}\}$, the objective function $\boldsymbol{\Theta}_{1+t}$ evaluated at the next point $\boldsymbol{\gamma}_{t+1}$ and $\boldsymbol{\Theta}_{1:t}$ are jointly Gaussian and the posterior distribution for $\boldsymbol{\Theta}_{1+t}$ is as follows [37]:

$$\boldsymbol{\Theta}_{1+t}|\mathbf{D}_{1:t} \sim \mathcal{N}\left(\mu(\boldsymbol{\gamma}_{t+1}),\sigma^2(\boldsymbol{\gamma}_{t+1})\right) \tag{13}$$

where

$$\mu(\boldsymbol{\gamma}_{t+1}) = \boldsymbol{k}^T\mathbf{K}^{-1}\boldsymbol{\Theta}_{1:t} \qquad \sigma^2(\boldsymbol{\gamma}_{t+1}) = k(\boldsymbol{\gamma}_{t+1},\boldsymbol{\gamma}_{t+1}) - \boldsymbol{k}^T\mathbf{K}^{-1}\boldsymbol{k} \tag{14}$$

$$\boldsymbol{k} = [\kappa(\boldsymbol{\gamma}_{t+1},\boldsymbol{\gamma}_1) \quad \kappa(\boldsymbol{\gamma}_{t+1},\boldsymbol{\gamma}_1) \quad \cdots \quad \kappa(\boldsymbol{\gamma}_{t+1},\boldsymbol{\gamma}_t)] \tag{15}$$

In Bayesian optimization, the point considered for the next evaluation $\boldsymbol{\gamma}_{t+1}$ corresponds to the maximum of an acquisition function. The peak in the acquisition function corresponds to either low predicted $\Theta(\boldsymbol{\gamma}_{t+1})$ value or high uncertainty in the prediction. The acquisition function used in this study is the Expected Improvement (EI). Details regarding the Expected Improvement can be found in [38]. Afterward, $\Theta(\boldsymbol{\gamma}_{t+1})$ is estimated, and the pair is then augmented to the original data. The GP is then updated, and the process is repeated until stopping criteria is achieved. Algorithm 1 shown below summarizes the steps for reconstructing the PDF using KDME with fractional moments method and Bayesian optimization. Additionally, a summary describing the Bayesian-optimized KDME novelty detection framework is presented in Fig. 1.

## Algorithm 1

**Input**: Sample points $\boldsymbol{x}$, number of sample points ns, PDF bounds $[x_{min},x_{max}]$, number of evaluation points $N$, number of order moments M, fractional powers bounds $[0, \gamma_{max}]$, bandwidth h $\in [0,\Delta x]$

**Output**: KDME-PDF $f_{KDME}$

1- set $z = (x - x_{min})/(x_{max} - x_{min})$
2- Select $N$ and calculate $x_{eval,i}$ $z_{eval,i}$ for $i = 1$ to $N$.
3- Perform Bayesian optimization using Eq. 10 as objective function and return $\gamma_{opt}$ under the constraint of $\gamma_k \in [0, \gamma_{max}]$
4- If multiple values in $\gamma_{opt}$ are close to $\gamma_{max}$, increase $\gamma_{max}$ and repeat step 3.
5- Estimate $\lambda$ (Eq. 7 and 8)
6- For $i = 1$ to $N$: calculate $p_i^{ME}$ using Eq. 6
7- For $j = 1$ to $n_s$: calculate $f_{KDME}(x_j)$ using Eq. 9

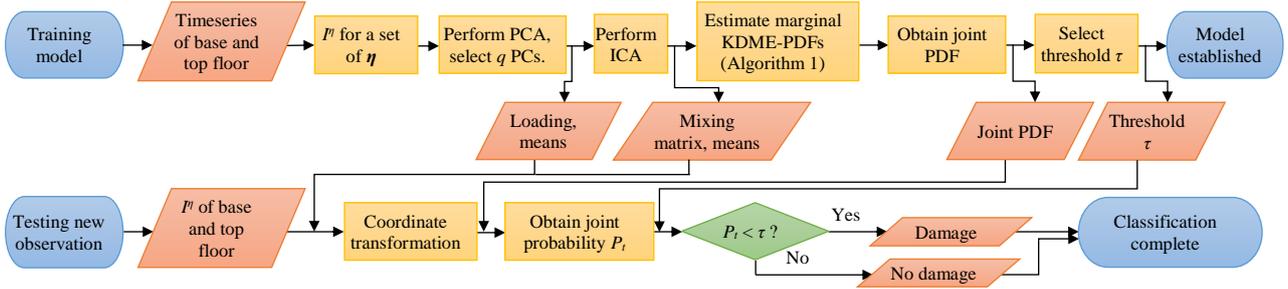

Figure 1 - Bayesian-optimized KDME novelty detection.

## 4. Numerical case study: a three-story reinforced concrete frame building

### 4.1 Model description

In this case study, the application of the proposed framework on a 3D RC moment frames building is shown. The frames are single-bay frames spanning 6.096 meters (20 ft) in each horizontal direction. The story height is 3.66 meters (12 ft). A schematic drawing of the structure showing cross-sections details is shown in Fig. 2(a). The structure is modeled using OpenSees [39]. All beams and columns are modeled as nonlinear beam-column elements with FiberSection objects. Additionally, P-delta effects are considered in the analysis. Environmental effects are incorporated in the form of varying temperatures accompanied by the temperature-dependent elastic modulus of the materials. The material models utilized are Concrete02 and Steel02. However, to simulate environmental variations effects, the elastic modulus for both concrete and steel are temperature-dependent since temperature variations effects dominate the SHM applications compared to other forms of environmental variations [40, 41]. The elastic modulus of steel is given by the following third-order polynomial [42]:

$$E_s(\tau) = 206 - 0.04326\tau - 3.502 \times 10^{-5}\tau^2 - 6.592 \times 10^{-8}\tau^3 \qquad (16)$$

where $\tau$ is the temperature in Celsius and $E_s(\tau)$ is the elastic modulus in GPa. For the compressive strength of concrete, the following is a proposed relation of the compressive strength and the temperature in Celsius valid for temperatures less than $100°C$ [43]:

$$f'_{c,\tau}(\tau) = f'_c[1.0 - 0.003125(\tau - 20)] \qquad (17)$$

where $f'_c$ is the compressive strength of concrete at 20°C. The concrete elastic modulus can be estimated using the compressive strength as per ACI 318-14 [44]. Finally, the concrete compressive strength at 20°C is 28 MPa while the steel yield strength is 460 MPa.

### 4.2 Data generation

The recording of the acceleration time-series at the base and the top floor in the two horizontal directions is done while the structure is under ambient vibration for both training and testing cases. The ambient vibration is simulated as white Gaussian noise with zero mean, 1E-4 g standard deviation and a sampling rate of 100Hz. For training data, the white noise is applied for 2880 minutes (48 hours). For the testing data, however, it is applied for 10 minutes after each earthquake simulation. The records are split into 60-second segments, resulting in a total of 2,880 training observations and 10 testing points per simulation. For the testing set, the structure is first subjected to an earthquake [51]. The 100 ground motions are selected from PEER NGA2-West database such that the magnitude is larger than 6.5, the Joyner-Boore distance is less than 30 km, and the soil class is D. A uniform scaling factor of 2.5 is applied to all ground motions [52]. To simulate temperature variations, the building is assumed to be located in San Diego, CA and the daily temperature records from the National Weather Service (NWS) online databases for San Diego, CA [45] are used. For every 10 minutes in the training data, a temperature value is randomly sampled from the daily temperature values of the year 2017 in a manner that ensures the inclusion of all seasons and different daily conditions. The same is done for each testing simulation however the temperature records used for testing are for the year 2018 (the subsequent year of training data acquisition year).

### 4.3 Damage definition and evaluation metrics

Defining the damage is essential for labeling the testing points. It is expected that the dominant failure mode is the flexural failure of columns (span-depth ratio > 4.0). Following ASCE 41-13 [46], damage at a story occurs if the plastic rotations of the columns at that particular story exceed the acceptance criteria of the *Immediate Occupancy* (IO) performance level, which is 0.5%. The plastic rotation can be approximated as the inter-story drift ratio which is available for each simulation. The structure is considered damaged if any inter-story drift ratio exceeds 0.5%. Otherwise, the structure is considered undamaged. Based on this criterion, there are 22 undamaged cases and 78 damaged cases. Fig. 2(b) shows a plot of peak story drift ratios for all testing cases.

The assessment is done based on 4 measures: accuracy, precision, recall, and F1 score, they are defined as follows:

$$Accuracy = \frac{TN+TP}{TN+TP+FN+FP} \times 100 \qquad Recall = \frac{TP}{TP+FN} \times 100$$

$$Precision = \frac{TP}{TP+FP} \times 100 \qquad F1-score = 2 \times \frac{Precision \times Recall}{Precision + Recall} \qquad (18)$$

where $TN$ is the number of true negatives, $TP$ is the number of true positives, $FP$ is the number of false positives (Type I errors), and $FN$ is the number of false negatives (Type II errors). It should be noted that in the case of the current SHM application, the emphasis should be given to recall because while false positives may waste time and/or resources, false negatives can be catastrophic.

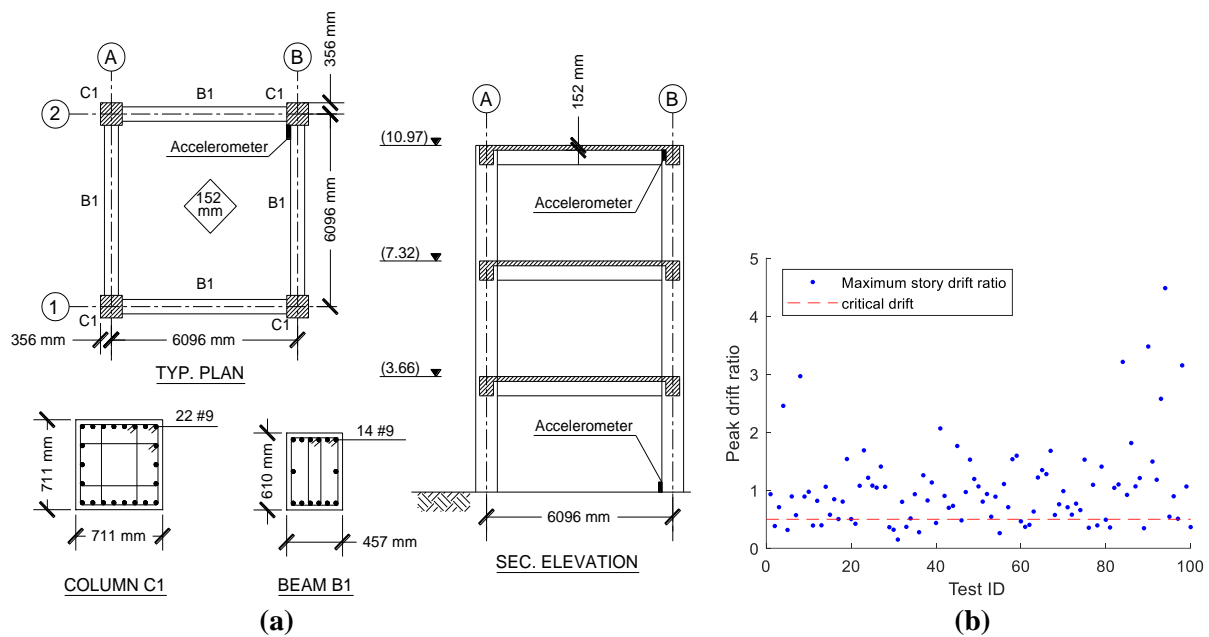

Figure 2 - (a) Structure details; (b) Peak story-drift ratios for testing simulations.

### 4.4 Results

Following the summary presented in Fig. 1, the damage-sensitive feature comprises the cumulative intensity measures of the acceleration time-series of the base and the top floor. The 400-dimensional feature vector shown in Eq. (2) is used. Min-max normalization with [0,1] bounds is done for each of the 400 features before performing PCA. Novelty detection models for 1 to 6 principal components are established and models are designated as PC# with # indicating the number of selected principal components. After obtaining the independent components using the robust ICA, Bayesian-optimized KDME is employed in reconstructing the marginal PDFs. Following extreme value theory, block-minima is applied to extract extreme joint probabilities of the training data with a window size of 30 points and the median of the extreme values is used as a threshold for each model. In the novelty test for testing cases, a voting system is put into effect for the observations generated from the same simulation case to decide on structure condition.

The testing results for all models in terms of evaluation metrics are summarized in Table 1. Accuracy of 94% is achieved for the best model (PC4) which is accompanied by 92.3% recall indicating the success of the Bayesian-optimized KDME novelty detection approach. Models with a low number of principal components performed poorly (e.g., PC1, PC2) which can be attributed to the lower model complexity and the low ratio of explained variance for the selected principal components. The explained variance ratios for these models are 27% and 48%, respectively. On the other hand, the best model has an explained variance ratio of 82%. It is also observed that with the increase in model complexity (e.g., PC6), the accuracy starts decreasing indicating that such models can be subject to overfitting if not enough data is available. Excluding the first three models with a low explained variance ratio, the mean accuracy for the remaining models becomes 92% bringing the conclusion that as long as the model is not too simple (e.g., 1 or 2 dimensional), a good model can be attained. The median joint probabilities of each simulation are plotted against the measured peak drift ratios for the best model and are shown in Fig. 3. The estimated for this model threshold is 6.7E-14.

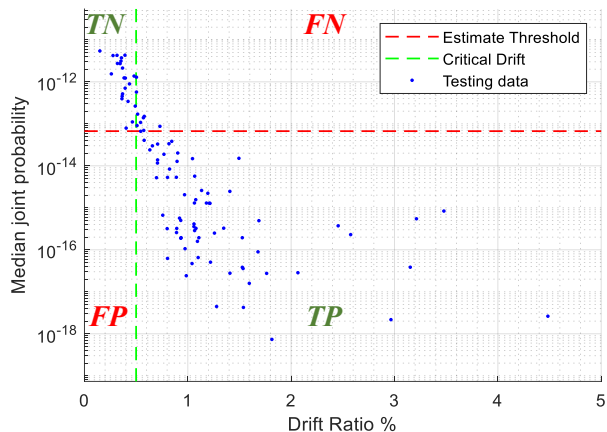

Figure 3 - Joint probabilities of testing observations vs drift ratio (PC4).

Table 1 – The evaluation metrics for test results.

| ID | FP | FN | Accuracy | Recall | Precision | F1 |
|----|-----|-----|----------|--------|-----------|-------|
| PC1 | 0 | 77 | 0.23 | 0.013 | 1.0 | 0.025 |
| PC2 | 0 | 78 | 0.22 | 0 | - | - |
| PC3 | 0 | 24 | 0.76 | 0.692 | 1.0 | 0.818 |
| PC4 | 0 | 6 | 0.94 | 0.923 | 1.0 | 0.96 |
| PC5 | 0 | 8 | 0.92 | 0.897 | 1.0 | 0.946 |
| PC6 | 0 | 10 | 0.90 | 0.872 | 1.0 | 0.932 |

## 5. Conclusions

In this study, a nonparametric unsupervised learning framework for structural health monitoring is proposed. The framework considers the cumulative intensity measure as a damage-sensitive feature for unsupervised learning. Additionally, PCA is implemented for selecting a few principal components of the feature vector while ICA transfer those components to the independent components space. The Bayesian-optimized KDME method is leveraged in PDF reconstruction allowing for accurate probability estimations for new observations, and consequently, establishing a robust SHM system for damage detection. The proposed approach is evaluated in a numerical case study of a three-story RC structure. In the case study, environmental variation effects, mainly temperature effects, are considered in testing the proposed approach. The following conclusions are made:

1- Benefiting solely from field-obtained normal condition data, unsupervised learning is an excellent choice as a data-driven SHM statistical learning method compared to supervised learning which, on the other hand, may require sophisticated numerical models for data generation.

2- The cumulative intensity measure $I^\eta$ is an effective damage-sensitive feature for unsupervised learning SHM methods. It can be suited for this task by performing PCA on a set of cumulative intensity measures with different values of the hyperparameter $\eta$ instead of tuning the hyperparameter in a supervised learning approach.

3- Introducing Bayesian optimization to the hyperparameter tuning of the fractional moment-orders in the KDME method greatly enhanced the PDF estimation process. Bayesian optimization is especially beneficial to the KDME method due to its improved performance in optimizing non-convex objective functions compared to other conventional optimization methods.

4- The implementation of ICA extends the KDME approach into the multivariate space paving way for KDME approach to be implemented in SHM where the feature vectors are typically multi-dimensional.

5- The proposed SHM approach yielded successful results when applied to the case study achieving damage detection accuracies up to 94% and up to 92.3% damaged cases recall.

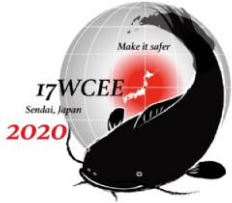

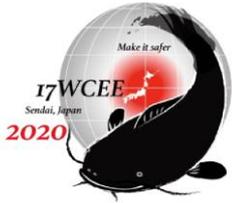